# Fine-grained Finger Gesture Recognition Using WiFi Signals

Sheng Tan[1], Jie Yang[1], Yingying Chen[2]

[1]Florida State University; [2]Rutgers University

**Abstract**—Gesture recognition has become increasingly important in human-computer interaction and can support different applications such as smart home, VR, and gaming. Traditional approaches usually rely on dedicated sensors that are worn by the user or cameras that require line of sight. In this paper, we present fine-grained finger gesture recognition by using commodity WiFi without requiring user to wear any sensors. Our system takes advantages of the fine-grained Channel State Information available from commodity WiFi devices and the prevalence of WiFi network infrastructures. It senses and identifies subtle movements of finger gestures by examining the unique patterns exhibited in the detailed CSI. We devise environmental noise removal mechanism to mitigate the effect of signal dynamic due to the environment changes. Moreover, we propose to capture the intrinsic gesture behavior to deal with individual diversity and gesture inconsistency. Lastly, we utilize multiple WiFi links and larger bandwidth at 5GHz to achieve finger gesture recognition under multi-user scenario. Our experimental evaluation in different environments demonstrates that our system can achieve over 90% recognition accuracy and is robust to both environment changes and individual diversity. Results also show that our system can provide accurate gesture recognition under different scenarios.

**Index Terms**—WiFi, channel state information, finger gesture

## 1 INTRODUCTION

IN recent years, gesture recognition is gaining increasing importance in human-computer interaction (HCI). Comparing to traditional techniques using peripheral devices such as mouse or keyboard, gesture-based interaction serves as a more convenient and natural means for users to interact with computers. Gesture made with fingers is particularly crucial as our HCI bandwidth is the highest there due to finger dexterity [2]. Recognizing finger gesture is also extremely compelling for interacting with mobile and wearable devices and performing finger control in emerging applications, such as smart home, virtual reality, and mobile gaming. Google's Soli radar chip [3], for example, is recently developed for the wearables to recognize finger gestures.

Prior work in gesture recognition mainly relies on pre-installed depth and infrared cameras (e.g., kinect, leap motion) [4], [5], [6], [7], [8] or dedicated sensors (e.g., RFID, gloves, motion sensors) that are worn by the user [9], [10], [11], [12], [13], [14], [15], [16]. These approaches however require significant deployment overhead and incur non-negligible cost. In addition, the camera-based solution cannot work in non-line-of-sight (NLOS) scenarios. Recently, Radio Frequency (RF) based gesture recognition using either specialized [17], [18], [19], [20] or commodity devices [21], [22], [23], [24] have drawn considerable attention as they don't require users to wear any physical sensors and can work under NLOS scenarios. These systems however only provide coarse-grained gesture recognition such as body activities [17], [18], [19], [21], [22], [25] or hand movements [26], [27], [28], [23]. While WiKey

[24] and the system proposed by Chen *et al.* [20] can recognize specific finger movements of typing, WiKey [24] requires the WiFi packets to be transmitted at outrageously high rate of 2500 pack- ets/second and is very sensitive to environmental changes, and the system [20] relies on specialized software-defined radio to extract radio wave features that are not reported in commodity RF device. Such limitations and the high infrastructure cost make these methods hard to deploy for gesture recognition in a practical and user friendly system.

In this paper, we demonstrate that the commodity WiFi can be exploited for fine-grained finger gesture recognition which is both easily deployable and low-cost. Our proposed system takes advantages of the fine-grained wireless channel measurement of Channel State Information (CSI) and the prevalence of WiFi network infrastructure. First, the detailed physical layer measurement of CSI is internally tracked by IEEE 802.11 MIMO and is readily available in commodity WiFi devices. Such fine-grained CSI is able to detect the minute environment changes that alter signal propagation and multipath. It is thus capable of capturing the subtle movements of fingers to provide fine-grained gesture recognition. Leveraging detailed CSI to recognize gestures doesn't require users to wear any sensors and can work under both LOS and NLOS scenarios. Second, the prevalence of WiFi network infrastructures enables the proposed system to reuse existing WiFi devices and networks without requiring dedicated or specialized hardware. The system could reuse existing WiFi signals, for example, the beacon signals of WiFi networks, to perform finger gesture recognition. Reusing existing WiFi infrastructures not only advances and extends the applications that could be supported by WiFi networks but also enables easy and large-scale deployment of the proposed system due to the proliferation of WiFi devices and networks [29].

In particular, our system, WiFinger, utilizes commodity WiFi devices to recognize finger gestures by examining the unique patterns exhibited in the detailed CSI. Accurately discerning the finger gestures is challenging, because the multipath, shadowing, and fading components of signal could be dynamic due to the environment changes. For example, people walking around or moved furniture could change the multipath environment and affect the signal propagation. Such changes could also be sensed by the detailed CSI and may distort the CSI pattern of the finger gesture. Moreover, there exists individual diversity of each user such as the finger length and movement speed. Even for the same user, the same finger gesture could be slightly different from time to time due to the lack of consistency. Furthermore, current commodity WiFi based gesture recognition systems cannot work well under the multi-user scenario. This is due to the signal reflections from different users are mixed at the receiver end and existing systems cannot separate the signal components that are mainly affected by individual user. Recently, system like WiMU [30] try to resolve this issue by finding the possible combination of different known gestures through exhaustive search. But such an approach is unable to work when one or more users perform gestures that is unknown to the system. This greatly limits its applicable scenarios.

To handle environmental changes, we propose an environmental noise removal mechanism which employs multipath mitigation and wavelet based denoising to filter out the environmental noises while trying to keep the CSI patterns resulted from the finger gestures. In particular, the multipath mitigation removes the signal components that arrive at the receiver through longer multipath propagation which are more likely affected by the changed environments, while the wavelet based denoising is used to further remove the high frequency noises while trying to keep sufficient details of CSI pattern for differentiating similar gestures. To deal with the individual diversity and gesture inconsistency, we propose to identify the principal components of the CSI pattern and to choose critical subcarriers that are sensitive to finger gesture for accurate gesture recognition. Specifically, the principal component identification exploits the idea of the intrinsic gesture behavior of the user [31] and extracts the gesture components which are invariant across the same set of finger gestures that one user performed.

To resolve multi-user gesture recognition issue, our system utilize multiple WiFi links and the larger bandwidth at 5GHz. The motivation of utilizing larger bandwidth (over 600MHz) at 5GHz band is to provide a higher distance resolution at around 0.3 meters. Such resolution is sufficient to achieve fine-grained gesture recognition when multiple users are performing gestures simultaneously within a typical indoor environment. On the other hand, 2.4GHz band with much smaller bandwidth (less than 100MHz) can only provide distance resolution at around 2 meters. There are existing WiFi devices such as laptops, IoT devices and APs in the smart home environment which can be further utilized to form multiple transmission links. The received signal from those links can be used to capture the RF signal propagation path change within the environment. Here we use power delay profile to quantify such change which gives the power strength of the received signal as a function of propagation delay. By subtracting the power delay profile under multi-user scenario (i.e., when multiple users perform finger gestures simultaneously) from the one under static environment (i.e., when there is no human presence), we can obtain the signal reflections that only affected by the finger motions of different users. Next, we separate the signal reflections from previous step into single user reflections profile which corresponds to each individual user. Then, we can achieve multi-user finger gesture recognition by analyzing the signal reflection components of each user at different transmission links. It is done by reconstructing the single user signal reflections as if there is only one user performing the finger gesture within the environment.

We experimentally evaluate WiFinger in both office and home environments with typical finger gestures including zoom in/out, circle left/right, swipe left/right, and flip up/down. Result shows that our system achieves overall accuracy over 93% and is robust to both environment changes and individual diversity. It also shows that our system can work with WiFi beacon signals and provides accurate gesture recognition under multi-user as well as NLOS scenario. The contributions of our work are summarized as follows:

· We show that the commodity WiFi can be reused to capture subtle changes of finger movements for fine-grained gesture recognition. Such approach doesn't require any dedicated or specialized devices and can work under NLOS scenarios.

· We devise environmental noise removal mechanism to mitigate the effect of the environment changes. Such a method enables the WiFinger's robustness to various environmental interference such as people walking around and furniture changes.

· We exploit the principal component of the CSI pattern and select critical subcarriers for accurate gesture recognition. The principal component extraction makes our system resilient to individual diversity and gesture inconsistency.

· We leverage all the available channels at 5GHz band from multiple transmission links to derive fine-grained power delay profile and separate the signal reflections from different users within the same multipath environment.

· We conduct extensive experiments in both office and home environments with multiple participants under various conditions. The results show that WiFinger achieves over 90% recognition accuracy and can work with existing WiFi beacon traffic, NLOS, and multi-user scenarios.

## 2 RELATED WORK

In general, the approaches for gesture recognition can be divided into three categories: wearable sensor based, camera based, and RF signal based.

**Wearable sensor based.** Many research efforts have been done by using dedicated sensors worn by users' hand for gesture recognition. For example, Risq [13] utilizes inertial sensors on a wristband to recognize smoking gestures.

Nelson *et al.* [10] developed a system using multi-sensor glove to recognize paralysis patients' gestures. Applications like text input using hand gestures also attract many attentions. PhonePoint Pen [9] for example recognizes human hand writing by holding mobile phone in hands. RF-IDraw [11] tracks hand or finger movements by attaching RFID to user's fingers. Other wearable devices such as smartwatch [14] or wearable ring [12], [16] can also be used to enable text input recognition by hand movements. These methods however all require user to wear physical sensors.
**Camera based.** Early works [7], [8] laid solid foundation for gesture recognition using dedicated cameras. Recent advancement in imaging technology enables depth or infrared cameras for gesture recognition, including the ones used in Microsoft Kinect [4], Leap Motion [5] and WiiU [6]. Although they do not require user to wear any sensors, they rely on dedicated hardware which incurs non-negligible installation overhead, and only work under LOS scenario.

**RF signal based.** The RF signal based methods are most related to our work. Without requiring user to wear any physical sensors, they can sense user motion under both LOS and NLOS scenarios [45-53]. By using specialized devices, systems like WiSee [17] WiTrack [19] and Wi-Vi [18] are able to track large scale movements. AllSee [26] and the system proposed by Chen *et al.* [20] are capable of tracking hand movements and even the finger movements of typing. Those systems however all rely on specialized hardware. Although the systems (e.g., E-eyes and CARM [21], [22]) use commodity WiFi devices, they can only identify large scale human activities or hand movements. While WiKey [24] can recognize finger typing motions, it requires the packets to be transmitted at outrageously high rate.

Much research has been done that attempt to resolve the issue of multi-user compatibility in commodity WiFi based gesture recognition systems. WiMU [30] can achieve multi-user gesture recognition by exhaustive search and compare of different known gestures combination to the collected samples. The proposed system can only work under the scenario where system has knowledge of all activities/gestures that can be performed by users. It cannot work when one or more users perform activities/gestures that are unknown to the system. Systems like CrossSense [32] and EI [33] leverage deep learning techniques to achieve better activity recognition performance under multi-user scenario when comparing with state-of-the-art systems. However, those approaches cannot work well without large number of training samples and require constant system update once the multipath environment changes.

Recently, many research have been done to achieve CSI-based gesture recognition adopting machine learning or deep learning models. Work like SignFi [34] focuses on recognizing large collection of sign language gestures, which involve the head, arm, hand, and finger gestures. It can achieve high accuracy utilizing CNN based classification algorithms. Wi-Multi proposed by Feng *et al.* [35] achieves large-scale gesture recognition utilizing either SVM or DTW depends on if a sufficient number of collected samples are available. Yang *et al.* [36] proposed a novel deep Siamese representation learning architecture for one-shot gesture recognition using CSI. Such system can achieve good performance when dealing with environmental dynamics and in-dividual heterogeneity. Widar 3.0 [37] achieve cross-domain hand gesture recognition using domain-independent feature and GRU model. WiCAR [38] utilizes multi-adversarial domain adaptation model to achieve in-car gesture recognition. However, those systems mainly focus on addressing the large-scale gesture/activity problem and only work well under single user scenario.

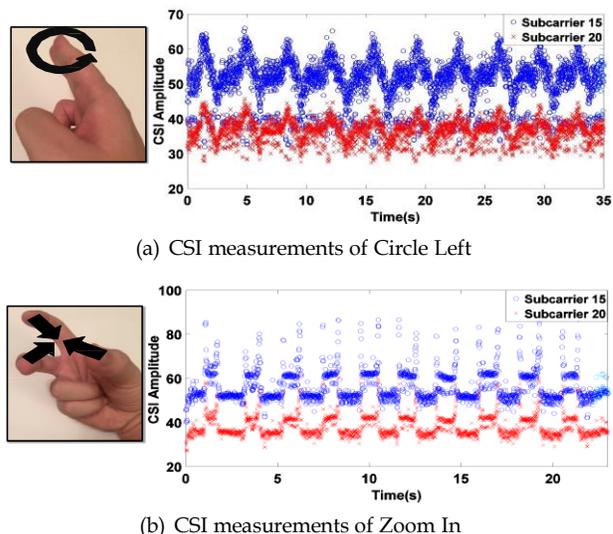

(a) CSI measurements of Circle Left

(b) CSI measurements of Zoom In

**Fig. 1:** Illustration of CSI measurements for two different finger gestures.

Comparing to the existing approaches discussed above, our proposed system can provide fine-grained finger gesture recognition using commodity WiFi while achieve multi-user compatibility. It is a software-only solution, which is both easily deployable and low-cost.

## 3 System Design

### 3.1 Preliminaries and Challenges

WiFi has been evolving from providing laptop connectivity to connecting all kinds of mobile and smart devices with higher speed and more advanced technologies. It has resulted in the prevalence of WiFi devices and ubiquitous coverage of WiFi network, which provides the opportunity to extend WiFi's capabilities beyond communication, particularly in sensing the physical environment. When the wireless signal propagates through space, any environment changes, either small scale or large scale, affect the received wireless signal, which is commonly known as shadowing and small-scale fading. With measurable changes in the received signals, activities in the physical environment could be potentially inferred. In particular, the 802.11a/g/n/ac employs OFDM technology which partitions the relatively wideband 20MHz channel into 52 subcarriers and provides detailed channel state information (CSI) of each subcarrier. The relative "narrow-band" subcarriers are very sensitive to the small movements in physical environment which results in the changes of CSI. On the contrary, the traditionally used received signal strength (RSS) is a coarse-grained information which provides averaged power in the received signal over the whole channel bandwidth and may not capture such changes. We are thus motivated to reuse existing WiFi infrastructure to sense and identify subtle movements of finger gestures by leveraging the detailed CSI provided by the commodity WiFi device.

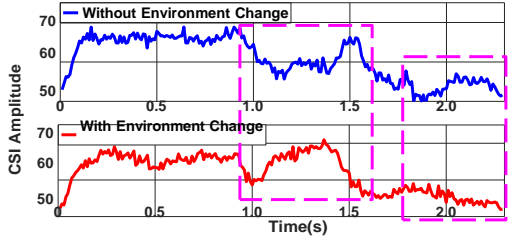

**Fig. 2: CSI patterns of Circle Left under different environments.**

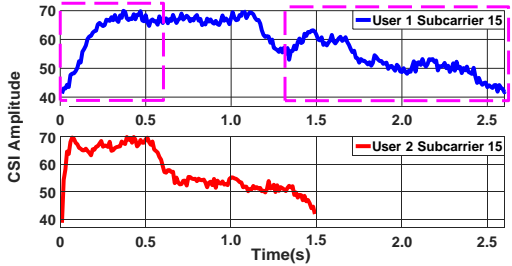

**Fig. 3: CSI pattern of Circle Left performed by different users.**

Figure 1 shows the CSI amplitude of two subcarriers(i.e., subcarrier 15 and 20) when a participant is repeatedly doing two finger gestures (circle left and zoom in) in front of the laptop. The CSI is extracted from the laptop that connected to a commercial AP in a 802.11n network. We observe that the CSI amplitude of those two subcarriers exhibits obvious periodic patterns and each of the finger gestures can be distinguished by its unique CSI pattern. This observation strongly indicates that the detailed CSI extracted from commodity WiFi could be analyzed for fine-grained figure gesture recognition.

Accurately discerning the finger gestures is however challenging because of the interferences from the surrounding environment. The interferences could come from the environment changes such as furniture change and people moving around. Such changes, for example a table/chair is moved to a different place or a person is walking around in the environment, alter the multipath environment which leads to construction or destruction (based on individual subcarrier phase shifts) effect in the combined signals at the receiver. Such effect could also be captured by the detailed CSI (due to subcarrier's relative "narrow-band" nature) and creates distortion of the CSI pattern. Figure 2 illustrates such CSI pattern distortion at one subcarrier when a participant is doing circle left gesture with and without environment changes. We observe that the CSI patterns in the dash windows are heavily distorted due to one person is walking around in the environment. Such distortion could significantly degrade the accuracy of gesture recognition.

Moreover, the finger gesture is subjected to individual diversity and gesture inconsistency. Different people may have different finger and hand size, movement pace, and habit to perform finger gestures. Even for the same person, she/he could perform the same gesture slightly different from time to time due to lack of gesture consistency. Figure 3 shows the captured CSI amplitude of one subcarrier for the same gesture performed by two different users. Although the shape of these two CSI traces exhibits certain similarity, the length and some details of the CSI pattern are very different due to different finger movement speed and gesture inconsistency. In particular, the second user perform finger gesture much faster than the first one, and the patterns at the head and tail of the CSI traces have clear difference. The individual diversity and gesture inconsistency thus could seriously affect the robustness of the recognition system.

Lastly, existing RF-based gesture recognition systems using commodity WiFi are mainly designed for and work with the single user scenario. Their performance suffer from severe degradation when multiple users are performing gesture simultaneously within the same environment. This is because the signal reflections captured at the receiver end are the mixture of multipath components from different users. Existing system cannot disentangle the signal reflections mainly affected by individual user. System like WiMU [30] attempts to resolve this issue by exhaustive search all the possible combination of different gestures that are known to the system. However, such an approach has limited applicable scenarios because it cannot work when one or more users perform unknown gestures.

### 3.2 Design Goals

To accurately recognize the fine-grained finger gestures by using the detailed CSI from a single commodity WiFi device, the design and implementation of our system involve a number of challenges:

**Easily Deployable.** The system should be easily deployable on existing commodity WiFi without using any dedicated or specialized hardware or requiring users to wear any physical sensors. It should work with LOS/NLOS and multi-user scenarios. It also should work only utilize existing WiFi traffic or beacons at the deployed AP without dedicated user generated traffic.

**Robust to Environmental Change.** The interferences from the surrounding environment could dynamically change the detailed CSI. Our system should be able to provide accurate finger gesture recognition by mitigating interferences such as furniture change, people moving around, and body movements of the user.

**Resilient to Individual Diversity and Gesture Inconsistency.** Once the system is setup, it should be able to be used by different users without user-specific calibration. It thus should be resilient to both individual diversity and gesture variation due to lack of consistency.

**Compatible to Multi-user Scenarios.** The proposed system should be able to achieve comparable performance with respect to single user scenario when there are multiple users perform finger gestures simultaneously within the same environment.

### 3.3 System Overview

The basic idea of our system is to examine the unique pattern exhibited in the CSI measurements that extracted from commodity WiFi devices. Figure 4 shows the system flow. WiFinger takes the time-series CSI measurements extracted from commodity WiFi devices as input. It can reuse existing network traffic, such as WiFi beaconing signals, or system-generated periodic traffic (if network traffic is insufficient) to extract the detailed CSI for single user scenario. For the multi-user scenario, our system scans all the available channels at 5GHz band, where the transmitter send out probe packets to all the 5GHz channels successively within coherence time and multiple receivers obtain the CSI measurements from the received packet. In our system, the users first need to identify if it is single user or multi-user scenario which will then trigger channel scanning mechanism.

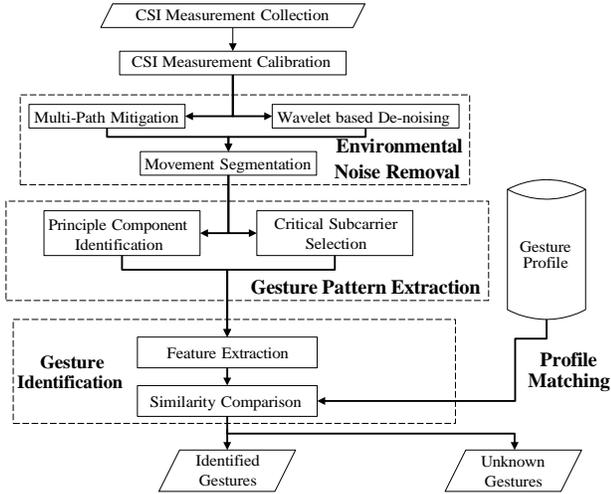

**Fig. 4: Overview of system flow.**

The obtained CSI measurements then go through *CSI Measurement Calibration* to mitigate phase error as well as amplitude error. For multi-user scenario, our system derives the fine-grained power delay profile leveraging non-uniform Discrete Fourier Transform (NDFT) from all the spliced channels. To obtain the single user signal reflection components, our system segment the derived power delay profile at each transmission link through a sliding window. Then, the fine-grained power delay profile of individual user is reconstructed as if he/she has performed gestures alone within the environment.

After each individual user's power delay profile is obtained, it goes through the *Environmental Noise Removal* and the *Gesture Pattern Extraction* component. Environmental noise removal encompasses two different techniques to address the challenge of environmental interferences. It first employs *Multipath Mitigation* to mitigate the interference stemmed from the environment changes such as moved furniture and/or people moving around. It then utilizes *Wavelet Based Denoising* to further remove the noise by decomposing signals into approximation coefficients and detail coefficients. A dynamic thresholding method is applied to the detail coefficients to remove the noisy components while keeping sufficient details of the CSI pattern. After that, the system performs movement segmentation to separate the CSI measurements to each finger gesture.

Next, our system performs gesture pattern extraction by utilizing *Principal Component Identification* and *Critical Subcarrier Selection*. Principal component identification is used to capture the intrinsic user gesture behaviors by identifying the CSI components which remain stable within the user's finger gestures. The identified principal components are usually invariant in the presence extensive variations in the user's gesture, and hence resilient to individual diversity and gesture inconsistency. Our system then uses critical subcarrier selection to choose the subcarriers that have high sensitivity to the subtle movements of fingers gestures for gesture recognition.

At last, our system recognizes gestures by going through *Gesture Identification* process. Our system first extract feature and then calculate the similarity of such feature with respect to each enrolled finger gesture profile using Muti-Dimensional Dynamic Time Warping (MD-DTW). The one with the profile in the library that has the highest and also sufficient similarity with the testing CSI pattern is then identified as the recognized gesture.

To construct gesture profile, our system could utilize either a supervised or semisupervised approach. For example, one user could perform each finger gesture several times offline and then label the corresponding extracted CSI pattern in the profile library. The system can also continuously monitoring user's gestures and identify multiple similar instances of CSI pattern without a matching profile. The user then could provide feedback to label such CSI pattern and deposit it to the profile library for subsequent gesture recognition. Moreover, the system could also use the semisupervised approach to update the CSI pattern when the gesture evolves to a slightly different version.

### 3.4 CSI Measurement Collection

Our system has the ability to extract CSI from the WiFi NIC which is the sampled version of the channel frequecy response when it is connected with 802.11n network. Specifically, for each of the 20MHz WiFi channel, the extracted CSI contains both amplitude and phase information for each of the 56 orthogonal frequency-division multiplexing (OFDM) subcarriers. In this work, for single user scenario, we use single WiFi device with its connected AP and one 20MHz channel to examine the unique patterns exhibited in the detailed CSI. Meanwhile, for multi-user scenario, we utilize multiple WiFi links and all the available 20MHz channels at 5GHz band to identify and extract the signal reflection from each individual user. To ensure WiFinger can probe through all the channels within the coherence time, we set the channel hopping delay to 0.2ms.

In particular, we denote the CSI measurements from all available channels at 5GHz band as:

$$\mathbf{csi_q} = [\mathbf{csi}_{1,q}, ..., \mathbf{csi}_{p,q}, ...], \tag{1}$$

where $p$ and $q$ denotes the $p^{th}$ channel at the $q^{th}$ receiver.

### 3.5 CSI Measurement Calibration

The raw CSI measurement extracted from previous step contain considerable distortions due to the hardware limitation of commodity WiFi NICs which mainly caused by clock unsynchronization of each transmission pair. Here, we achieve CSI calibration by adopting the correction approach proposed by previous work [39]. In particular, we average CSI measurements from several packets that collected within coherence time to mitigate amplitude error. For the phase error, it involves both constant component and linear component. For the constant component, we mitigate the error by selecting a reference channel and compensate the difference between such channel and all the other collected channels. For the linear component, we first average multiple CSI measurements at the same channel at individual receiver and then search for an optimized offset that can minimize the difference between all pairs of power delay profile derived from the channels collected within coherence time. After calibration, the CSI measurement at each transmitter and receiver pair can be represented as:

$$\mathbf{CSI_q} = [\mathbf{CSI}_{1,q}, ..., \mathbf{CSI}_{p,q}, ...], \quad (2)$$

Given the calibrated channel response CSI, the power delay profile $\mathbf{g}$ at given channel can be derived using IFFT:

$$\mathbf{g}_q = \sum_{r=1}^{R} a_r \delta(t - t_r), \quad (3)$$

where $r$ denotes the sequence number of total R multipath, $a_r$ and $t_r$ are the amplitude and signal propagation time delay of $r$th path, $\delta(t)$ is the Dirac delta function. For multiple user scenario, in order to derive a more fine-grained power delay profile, our system needs to stitch together the CSI measurements collected from all the channels at 5GHz after calibration. But all the available channels are unequally and non-contiguous spaced at 5GHz band due to the regulation issue. Particularly, all the available channels on the Intel 5300 NICs at 5GHz band are divided into three different segments: first segment from channel 36 to 64 (5.17GHz to 5.33GHz), second segment from channel 100 to 140 (5.49GHz to 5.71GHz) and third segment from channel 149 to 165 (5.735GHz to 5.835GHz). Instead of using simple IFFT which only works for evenly spaced channels, we propose to utilize inverse Non-uniform Discrete Fourier Transform (NDFT) which can be applied to non-uniformly spaced channels. Here we can formulate the inverse NDFT problem as following:

$$\min_{\mathbf{g}} ||\mathbf{CSI_q} - \mathbf{F}\,\mathbf{g}\,||_2^2, \quad (4)$$

where $\mathbf{g}$ represents the power delay profile we are trying to find and $\mathbf{F}$ is Fourier matrix. The goal is to search for an optimum solution of power delay profile that can minimize the difference between the Fourier Transformation of $\mathbf{g}$ and spliced CSI measurements from all available channels.

Because the search for power delay profile will yield non-linear and non-closed form results, we leverage the layout information of transmission links to select the optimum solution. Here, we assume currently all the signal propagation from each transmission pair has line-of-sight where the LoS propagation path has the largest power among all the multipath propagation. Thus, our system favors the derived power delay profile that has larger power at the LoS propagation among all the yield solutions. By leveraging inverse NDFT, we can further improve the resolution of the derived power delay profile.

After obtaining the fine-grained power delay profile, our system performs multi-user profile separation and reconstruction to acquire the signal reflections from each user. The basic idea is illustrated in Figure 5. Because the obtained power delay profile consists of signal reflections components from both the motions of multiple users and static environment, we first subtract the derived power delay profile at one of the WiFi links *linkT1R1* as shown in Figure 5 (b) by the profile under the static environment shown in Figure 5 (a) which is collected when there is no human presents within the environment. Our system can then determine the number of users by detecting how many major signal reflection components exist after subtraction. Next, we further segment the subtracted power delay profile representing only multiple user motion information as shown in Figure 5 (c) into single user reflections profile where each one contains the signal reflection components mainly caused by individual user. This is done by using a moving window based approach. In particular, the power differential between adjacent time points is accumulated within each sliding window and compare to a threshold to determine the duration of the individual user motion. Here, we empirically set the threshold to be 0.6. This process is repeated over all the power delay profiles derived from subtraction at each transmission link.

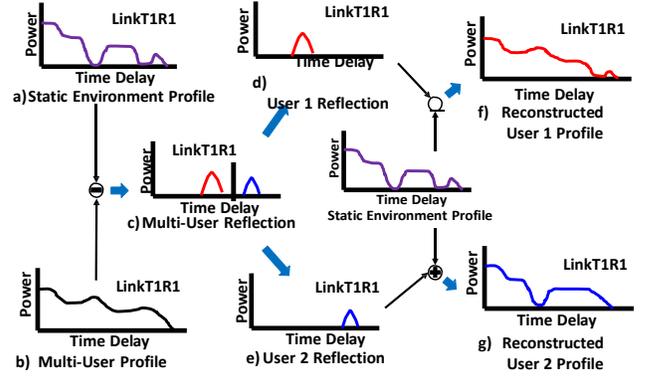

**Fig. 5: Illustration of multi-user profile separation and reconstruction.**

It is worth noting, when multiple users are at the same distance to one particular transmission link, the signal reflection from these users could overlap (i.e., signal reflections from different users share the similar propagation time delay) in the multi-user reflection profile at that link. Thus, simply utilizing power delay profile derived from a single link could not distinguish multiple users. Here, we propose to use multiple transmission links (e.g., 3 or more) to overcome this problem. Due to geometric relation between multiple transmission links (e.g., 3 or more), one or more transmission links could capture signal reflection from multiple users without overlapping. Therefore, we can further separate different users' profiles based on multiple transmission links that are not severely overlapping.

At last, the segmented power delay profile obtained from previous step as shown in Figure 5 (d) and Figure 5 (e) will go through individual profile construction to reconstruct the signal reflection profile dominated by individual user as if there is only one user performing gesture within the environment. This is done by combining the segmented power delay profile with the power delay profile collected under the static environment scenario. After this, we are able to derive the power delay profile shown in Figure 5 (f) and Figure 5 (g). Such reconstructed profiles contain both signal reflection mainly caused by the target user motion and the static environment. Meanwhile, the interference from the motion of other users has been mitigated.

### 3.6 Environmental Noise Removal

In this subsection, we present the details of two techniques the system used to mitigate the interferences from the surrounding environment: multipath mitigation and wavelet based denoising.

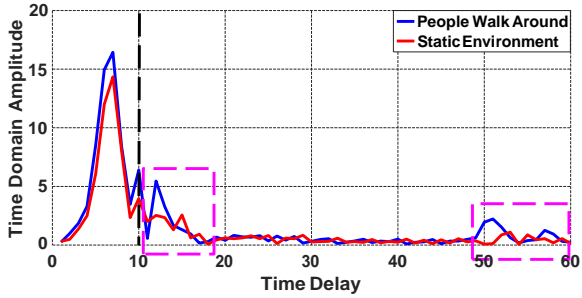

**Fig. 6: Power delay profile with and without environment change.**

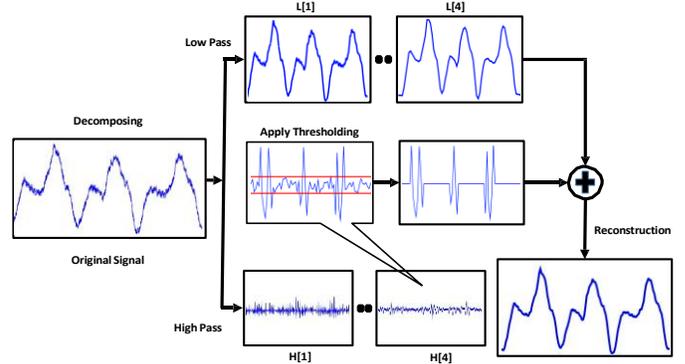

**Fig. 7: Illustration of Wavelet based denoising.**

### 3.6.1 Multi-Path Mitigation

Multipath mitigation aims to remove the signal components that arrive at the receiver through longer multipath propagation. As the environment changes such as a table/chair is moved to a different place or a person is walking around will reflect the wireless signal and also create additional multipath, these reflected signals via created multipath will distort the CSI pattern of a finger gesture. Removing these reflected signal components could mitigate such interferences and hence makes the system robust to the environment changes. In particular, the signal reflection via multipath usually has longer propagation delays before arriving at the receiver. By transferring the frequency domain CSI into time-domain power delay profile, we could remove the signal components that have longer delay to mitigate the effect of the changed multipath.

Figure 6 shows the power delay profile with 60-point IFFT for the same gesture (i.e., circle left) under the scenarios with and without people moving around in the environment. We observe that the signal components in these two dash windows have obvious difference due to the environment change. We thus remove the signal components with large time delay (i.e., the part on the right side of the dash line in Figure 6) to retain the CSI pattern of the finger gesture while mitigating the effect of the changed multipath environment. After removing the signal components with larger delay, we apply an FFT transformation to covert the trimmed profile to frequency domain CSI. Previous study shows general indoor environment has the maximum delay less than 500 ns [40]. We use this value as a baseline for removing the signal components with longer delay, shown as the dash line in Figure 6.

### 3.6.2 Wavelet Based Denoising

Wavelet based denoising is used to further remove the noises presented in the collected CSI measurements. These interferences could come from various sources such as the nearby electric devices and WiFi devices' inner noise. It is based on the Discrete Wavelet Transform (DWT) which analyzes the signal in both time and frequency domain and doesn't make any assumption about the nature of the signal. The DWT decomposes signals into approximation coefficients and detail coefficients. While the approximation coefficients describe the shape/trend of the signal which retain large scale characteristic of the CSI pattern, the detail coefficients capture the low-scale components which represent both high frequency noise and the fine details of the CSI pattern. As we are interested in removing the high frequency noise components while trying to keep sufficient details of CSI pattern for differentiating similar gestures, a dynamic thresholding is applied to the detail coefficients to remove the noisy components.

In particular, the wavelet based denoising includes three steps: decomposition, thresholding, and reconstruction. As shown in Figure 7, we first run the DWT based signal decomposition recursively by four levels with Symlet wavelet filter [41]. The DWT then yields both approximation coefficients $\alpha^J$ (with $J = 4$) and a sequence of detailed coefficients $\beta^1, \beta^2, ... \beta^{(J)}$. Each level of DWT coefficients are computed based on the following equations:

$$\alpha_k^{(J)} = (x_n, g_{n-2^J k})_n = \sum_{n \in Z} x_n\, g_{n-2^J k}, \; J \in Z \quad (5)$$

$$\beta_k^{(\pounds)} = (x_n, h_{n-2\pounds k})_n = \sum_{n \in Z} x_n\, h_{n-2\pounds k}, \; f \in \{1, 2, ..., J\} \quad (6)$$

where $x_n$ is the $n^{th}$ input point, $(.)$ is the dot product operation, and wavelet basis represents by two sets of discrete orthogonal functions $g$'s and $h$'s.

We then apply dynamic thresholding to each level of detail coefficients $\beta^1, \beta^2, ... \beta^{(J)}$ to remove their noisy components. Finally, by combining all the resulting coefficients (i.e., the approximation coefficients and the detail coefficients after noisy removal), we reconstruct the final denoised CSI measurements with the inverse DWT. The inverse DWT is given by following formula:

$$x_n = \sum_{k \in Z} \alpha_k^{(J)}\, g_{n-2^J k} + \sum_{\pounds=1}^{\cdot} \sum_{k \in Z} \beta_k^{(\pounds)}\, h_{n-2\pounds k} \quad (7)$$

The reconstructed measurements enable us to remove the noise components while keeping the detailed patterns. This could facilitate accurate gesture recognition, especially for those gestures with similar shape of CSI patterns.

The system requires a user to have a short static interval between gestures to serve as the sentinel signal. Our system then can identify the movement of a gesture by detecting the static interval. In detail, we accumulate the amplitude differential between adjacent time points within each sliding window. The accumulated value is then compared to a empirical threshold for determining the sentinel signal for CSI trace segmentation.

### 3.7 Gesture Pattern Extraction

We next detail the gesture pattern extraction component which is used to identify the principal components of CSI patterns and to choose critical subcarriers for accurate gesture recognition.

#### 3.7.1 Principal Component Identification

The principal component identification borrows the idea of the intrinsic gesture behavior of the user in signature verification [31]. In particular, the CSI measurements of each finger gesture could be divided into several gesture components. Due to the individual diversity and gesture inconsistency, only part of these components are invariant across the same set of finger gestures that one user performed. We refer such components as principal components which capture the intrinsic gesture behavior of the user. Our system thus extracts these principal components to facilitate gesture recognition for improving the resilience to individual diversity and gesture inconsistency.

To identify the principal components of the CSI pattern, we examine and compare multiple instances of the same gesture. In particular, our method takes two instances of CSI measurements and compares them to find the best alignment by calculating a cost matrix and discovering the lowest cost route. The resulted lowest cost route could be represented by a coupling sequence in which the direct matching samples denote the components without significant distortion between two instances. We thus incorporate these direct matching samples into a weight vector to represent the principal components of the finger gesture. We run this process repeatedly between different pairs of instances that are available during the profile construction phase, and then average over the resulting weight vectors to obtain the principal components of each finger gesture.

Following shows the details of the principal component identification algorithm. After environmental noise removal, we first interpolated CSI measurements of each gesture instance to a fixed length $L$. We then assume $\{c_i, 1 \leq i \leq N\}$ is a set of interpolated CSI measurements with the fixed length $L$ extracted from $N$ gesture instances. The weight vector derived from a pair of instances $c_i$ and $c_j$ can be described as: $w^{c_i, c_j}$ where $i \mathrel{/}= j$ and $1 \leq l \leq L$. We then use the coupling sequence which is the alignment between $c_i$ and $c_j$ to estimate the weight value. All the direct matching samples in the coupling sequence are considered as the principal component candidates which represent the consistent gesture segments between two CSI instances. We simply set 1 as the weight if it is a principal component, and assign a weight 0 otherwise. At last, we generalize the principal components by averaging the weight vectors over each pair of CSI instances. Each averaged weight value ranges from 0 to 1 indicating the consistency of the corresponding segment of the CSI measurements. And a larger weight value means the corresponding segment is more stable when performing finger gestures. Our system thus values the segments with higher averaged weights more significantly during the gesture identification procedure as they represent the intrinsic gesture behavior.

Figure 8 shows one example on the process of the principal component identification using two instances. We first calculate the cost matrix between these two instances, as shown in Figure 8(b). Based on the coupling sequence shown in the cost matrix, we identify these direct matching samples as the candidates of the principal components of the gesture. We then map these direct matching samples back to the CSI measurements of those two instances. The identified principal components in these two instances are highlighted as black color, as shown in Figure 8(c). During the gesture recognition phase, the principal components will be assigned with higher weights while the rest will be assigned with lower weights. The principal components identification thus could effectively capture the intrinsic gesture behaviors and improve the robustness of the system.

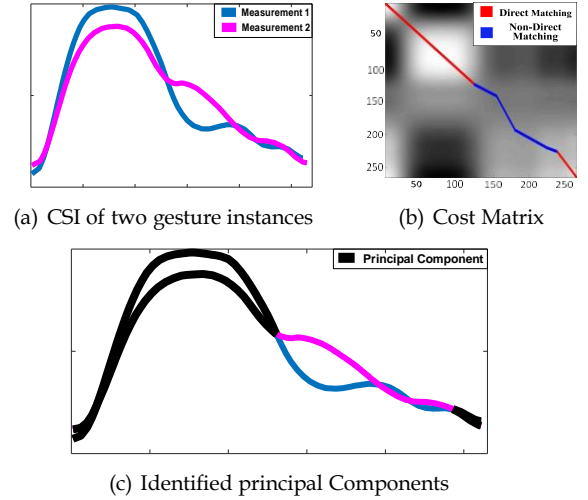

(a) CSI of two gesture instances    (b) Cost Matrix

(c) Identified principal Components

**Fig. 8: Illustration of principal component identification steps.**

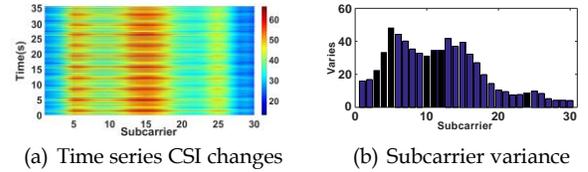

(a) Time series CSI changes    (b) Subcarrier variance

**Fig. 9: One example of subcarrier sensitivity to the finger gesture of Circle Left.**

#### 3.7.2 Critical Subcarrier Selection

Due to the frequency diversity, different subcarriers have different sensitivity to the subtle movements of finger gestures [42]. Figure 9 illustrates an example of time series CSI changes for 30 subcarriers when performing circle left gesture. We observe that the subcarriers with smaller indices are more sensitive to the circle left gesture, while the CSI from the higher subcarrier indices presents less changes. It is thus desire to assign more weights to these subcarriers with higher sensitivity for gesture recognition. Specifically, we calculate the variance of the CSI in a moving window in time series to quantify the sensitivity of the subcarriers to the finger gesture. By comparing to other work simply discard the subcarriers with less sensitivity, our approach enhances the impact of more sensitive subcarriers while preserving the effect of subcarriers with lower sensitivity.

### 3.8 Gesture Identification

To extract feature from finger gesture motions, we propose to use Doppler shift by leveraging short-term Fourier transform (STFT). Such an approach will compute the spectrogram which represents the time-frequency information of the given motion. In particular, we apply STFT to the CSI measurements obtained from previous step with a Gaussian window (i.e., < 0.1s) where we assume Doppler shift is invariant during that time period. Our system then extracts the energy based frequency contour of derived spectrogram. Specifically, the power level of given spectrogram is normalized into the same scale from 0 to 1. Then, a pre-defined power level band is chosen (i.e., between 0.90 and 0.95) and the centroid frequencies at such band is combined which resulted in one positive and one negative frequency contour. After that, the extracted energy-based frequency contour will be utilized for similarity comparison.

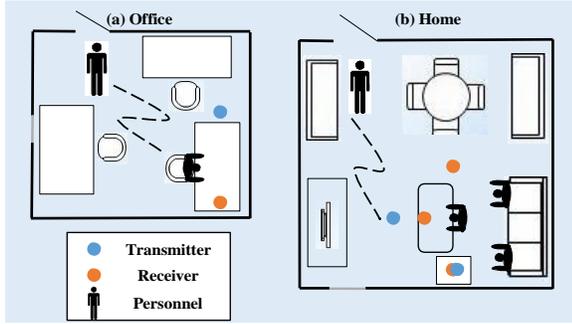

**Fig. 10: Illustration of experimental setup.**

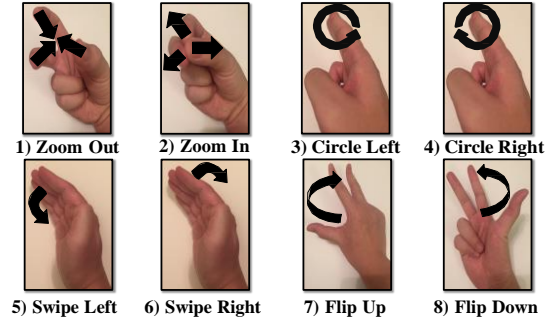

**Fig. 11: Illustration of eight finger gestures.**

To better facilitate similarity comparison, Muti-Dimensional Dynamic Time Warping (MD-DTW) [22] is utilized to achieve better alignment. MD-DTW allows us to overcome the problem of speed problem by focusing on shifts in the matching pattern. Thus it provides a robust metric for measuring the similarity. Specifically, we adopt the Euclidean distance to quantify the warping path. During gesture recognition, the extracted energy-based frequency contour is used as feature and MD-DTW is utilized to calculate the similarity between the testing instances and enrolled finger gesture profile. It is shown as:

$$d(s_a, t_b) = \sum_{u=1}^{U}(s_{a,u} - t_{b,u})^2 \tag{8}$$

where $S = s_1, s_2, ..., s_u$ and $T = t_1, t_2, ..., t_u$ are two CSI patterns where $U$ is the number of chosen subcarriers. The one with the highest and sufficient similarity in the gesture profile is identified as the recognized gesture whereas the one with insufficient similarity (lower than 0.75) to existing gesture is then identified as unknown gesture.

# 4 PERFORMANCE EVALUATION

In this section, we evaluate the performance of our WiFinger system using a commodity WiFi device in both office and home environments with multiple participants under various conditions.

## 4.1 Experimental Setup

### 4.1.1 Device and Network

For single user scenario, we conduct experiments with a single WiFi device (i.e., Dell LATITUDE E5540 Laptop) connected to a commercial wireless Access Point (LINKSYS E2500 N600 Wireless Router) in an 802.11n WiFi network.

The laptop runs Ubuntu 10.04 LTS and is equipped with an Intel WiFi Link 5300 for extracting CSI measurements [43]. The packet transmission rate is set to 20 pkts/s. We will discuss the impact of packet rate on overall recognition accuracy in Section 4.7. For each packet, we extract CSI for 30 subcarrier groups, which are evenly distributed in the 56 subcarriers of a 20MHz channel.

For multi-user scenario, we conduct experiments with one laptop as transmitter and three laptops as receiver. The setup of transmission links are shown on Figure 10(b). All laptops have the same software and NIC setup as stated in single user scenario. Both transmitter and receivers are in monitor mode and have the ability to send and receive packet at 5GHz band in an 802.11n network. The transmitter scans all the 24 available 20MHz channels at 5GHz of Intel 5300 NIC by sending packets at each channel to all receivers and each receiver will collect the packets of the particular channel during the scanning process. They can be further divided into three non-contiguous parts: from 5.18GHz to 5.32 GHz (i.e., the channels from 36 to 64), from 5.5GHz to 5.70GHz (i.e., the channels from 100 to 140) and from 5.73GHz to 5.83GHz (i.e., the channels from 149 to 165). As indicated in the previous work [44], typical indoor environment has the coherence time of several hundreds milliseconds. Thus, we set the channel hopping delay to 0.2ms as it allows us to collect multiple packets across all available channels within the coherence time. We then extract the CSI measurement contains 30 subcarriers which are equally distributed at a 20MHz channel for each packet.

### 4.1.2 Environments and Finger Gestures

We conduct experiments in both an office and an apartment environments with five participants. The experimental setup in these two environments are shown in Figure 10. The office has the size of about 9 ft by 9 ft with three tables and chairs, and some electronic devices inside, whereas the apartment is about 16 ft by 13 ft with regular living room furniture setup, such as dining table, book shelf, sofa, and TV. The office environment represents a more compact space filled with furniture, while the apartment setup describes a typical home environment with larger space. When the participant is performing the finger gesture, she/he is sitting on the sofa in the apartment environment and sitting in front of the table in office environment respectively. The AP and the laptop are placed at two sides of the sofa and table for single user scenario, as shown in Figure 10. For the multi-user experiments, two users are sitting on the sofa when performing the finger gesture under two user setup while an additional user is standing in front of the table under three

|  | Swipe Left | Swipe Right | Zoom In | Zoom Out | Circle Left | Circle Right | Flip Up | Flip Down |
|---|---|---|---|---|---|---|---|---|
| Swipe Left |  | 0.00 | 0.00 | 0.01 | 0.00 | 0.01 | 0.00 | 0.02 | 0.01 |
| Swipe Right | 0.00 |  | 0.02 | 0.00 | 0.01 | 0.02 | 0.00 | 0.01 | 0.01 |
| Zoom In | 0.00 | 0.03 |  | 0.00 | 0.00 | 0.02 | 0.00 | 0.02 |
| Zoom Out | 0.00 | 0.03 | 0.03 |  | 0.00 | 0.00 | 0.03 | 0.00 | 0.02 |
| Circle Left | 0.01 | 0.00 | 0.02 | 0.02 |  | 0.02 | 0.01 | 0.03 |
| Circle Right | 0.01 | 0.00 | 0.01 | 0.02 | 0.01 |  | 0.00 | 0.02 |
| Flip Up | 0.01 | 0.01 | 0.00 | 0.01 | 0.02 | 0.01 |  | 0.02 |
| Flip Down | 0.01 | 0.01 | 0.02 | 0.01 | 0.00 | 0.01 | 0.02 |  |

(a) Home environment

|  | Swipe Left | Swipe Right | Zoom In | Zoom Out | Circle Left | Circle Right | Flip Up | Flip Down |
|---|---|---|---|---|---|---|---|---|
| Swipe Left |  | 0.00 | 0.00 | 0.01 | 0.02 | 0.00 | 0.02 | 0.01 |
| Swipe Right | 0.00 |  | 0.00 | 0.02 | 0.00 | 0.03 | 0.01 | 0.02 |
| Zoom In | 0.00 | 0.02 |  | 0.00 | 0.00 | 0.03 | 0.00 | 0.02 |
| Zoom Out | 0.00 | 0.02 | 0.03 |  | 0.00 | 0.02 | 0.03 | 0.01 |
| Circle Left | 0.00 | 0.01 | 0.01 | 0.01 |  | 0.00 | 0.01 | 0.03 |
| Circle Right | 0.03 | 0.00 | 0.01 | 0.01 | 0.02 |  | 0.00 | 0.03 |
| Flip Up | 0.01 | 0.01 | 0.00 | 0.02 | 0.01 | 0.02 |  | 0.02 |
| Flip Down | 0.01 | 0.01 | 0.02 | 0.01 | 0.00 | 0.01 | 0.02 |  |

(b) Office environment

**Fig. 12: Confusion matrix of finger gesture recognition under different environments.**

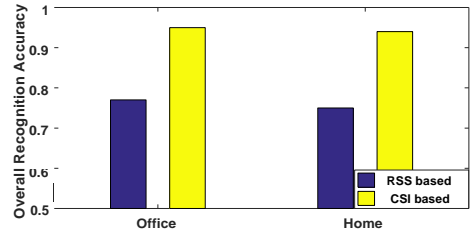

**Fig. 13: Performance comparison when using CSI and RSS.**

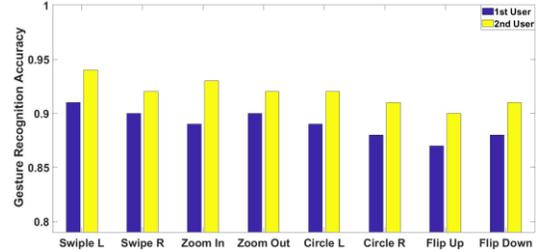

**Fig. 14: System performance under two-user scenario (7 ft. distance).**

user setup. The transmitter and receivers' positions under multi-user scenarios are marked respectively on Figure 10.

We evaluate the performance of our system with eight commonly used finger gestures including swipe left, swipe right, zoom in, room out, circle left, circle right, flip up, and flip down, as shown in Figure 11. These gestures are also widely used in current human-computer interaction systems such as Microsoft Kinect or Leap Motion. Each participant performs one gesture fifty times in office and apartment environments respectively. We use ten instances of each finger gesture to extract the CSI pattern for building the gesture profile. To test the robustness of our system to environment changes, we experiment with both furniture move and people walking around scenarios. In particular, when the participant is performing finger gesture, a second person is randomly walking around within the environment to create interference. Examples of the walking trajectories are shown in dash curve in Figure 10. For the furniture change, we move the chairs and tables from one place to another inside the room.

### 4.1.3 Metrics

We use both confusion matrix and recognition accuracy to evaluate the performance of our system.

**Confusion Matrix.** Each column represents the finger gesture that was classified by our system and each row shows finger gestures performed the user. Each cell in the confusion matrix represents the percentage of finger gesture in the row that was classified as the gestures in the column.

**Recognition Accuracy.** The percentage of the finger gestures correctly classified by our system.

### 4.2 Overall Performance

Figure 12 shows the confusion matrix of finger gesture recognition under both home and office environments. We observe that in both environments, our system achieves an overall recognition accuracy over 93% with the standard deviation at about 1.5%. By comparing the details of each finger gesture recognition in these two environments, we find that the recognition accuracy distribution are similar. In both environments, the swipe left and right have the highest recognition accuracy, whereas the flip up and down have the lowest accuracy. In particular, the swipe left achieves 96% and 95% accuracy in home and office environments respectively. This is possibly because of the relative larger finger movements involved in swipe left and right. Consequently, more finger movement details could be captured by CSI for differentiating from other similar finger gestures. The above results show that our system could provide high accuracy in recognizing finger gestures by using only a single WiFi device. The recognition accuracy could be further improved, for example, by using multiple available devices or the WiFi device equipped with multiple antennas.

We also compare the performance of using CSI to that of using RSS for finger gesture recognition. As RSS is the more sensitive to the physical movements when the transmitter and receiver are closer due to the log distance propagation, we place the WiFi device and the AP very close to each other (i.e., 3 ft) and compare the performance of CSI-based and RSS-based recognition in the same setup. Figure 13 illustrates the performance comparison of the overall recognition accuracy with each finger gesture tested for fifty times in each of these two environments. We observe that under the same setup, the CSI based method could achieve around 95% accuracy in both environments, whereas the RSS based method has only 76% recognition accuracy. It indicates that the detailed CSI could provide more fine-grained information than that of RSS, and can result in much better gesture recognition accuracy.

### 4.3 Multi-User Gesture Recognition Performance

We then evaluate the performance of our system in multi-user scenarios. The experiment is conducted in home environment where two users are performing finger gestures simultaneously. The results are shown in Figure 14. We observe that the overall recognition accuracy for two users are around 90% which is comparable to the single user scenario. Moreover, by comparing Figure 14 and Figure 12(a), we observe that the system performance does not have obvious degradation. This study demonstrates our system can achieve similar performance under multi-user scenarios. Furthermore, we can observe that our system achieves better recognition accuracy for user 2 with respect to user 1. Such phenomenon is likely caused by the placement of Wi-Fi receivers. With only three receivers for our current system setup, the received signal reflections are highly depending on the user locations with respect to each receiver. For example, when the distance between one of the users and all the receivers is shorter compare to the other users, the signal reflections from such user are stronger. Therefore, the derived power delay profile will contain more information of the gestures performed by this user which causes the recognition accuracy higher than all the other users. By adding more receivers or adjusting receiver placements can mitigate the phenomenon of recognition accuracy gap between different users.

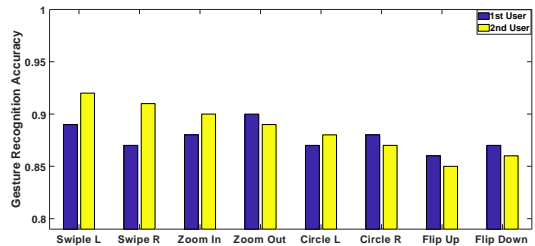

**Fig. 15: System performance under two-user scenario (5 ft. distance).**

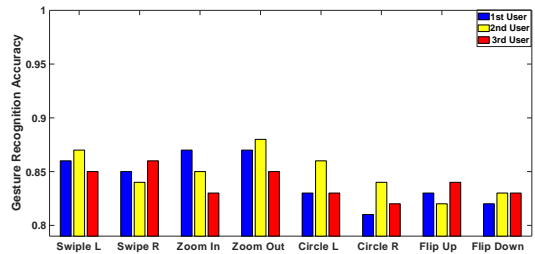

**Fig. 16: System performance under three-user scenario.**

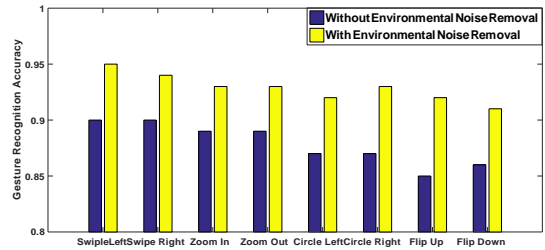

(a) Furniture change

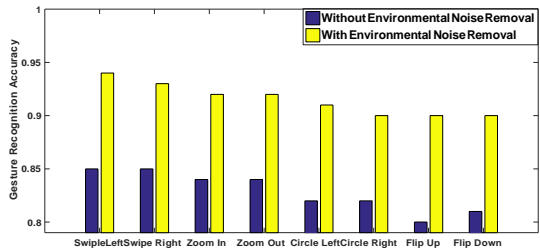

(b) People walking around

**Fig. 17: Recognition accuracy under the environment changes.**

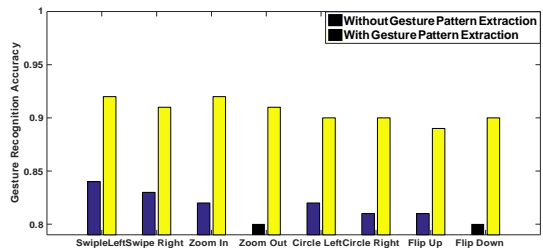

**Fig. 18: Recognition accuracy under individual diversity.**

As shown in Figure 15, we observe that the overall recognition accuracy for two users are over **88%** under the close proximity scenario where the distance between two users are 5 feet. Furthermore, by comparing Figure 15 and Figure 14 where the distance between two users are 7 feet, we observe similar system performance. This shows our system can achieve good performance even when multiple users are in close proximity.

We further evaluate the performance of WiFinger under multi-user scenario by allowing three different users perform finger gestures simultaneously. The user locations are illustrated in Figure 10, where the distance between each pair of users are 5 ft, 5 ft and 7 ft respectively. The system performance is shown in Figure 16. We can observe the overall recognition accuracy for three different users is around **85%**. This demonstrate our system has the ability to track more than two users simultaneously. We note that the recognition accuracy under three users are lower than that of the two users. This is because the signal reflections can be easier to separate under two-user scenarios compare to three-user scenario. Thus, a higher density of wireless links in the environment could potentially help to improve the recognition accuracy.

### 4.4 Impact of Environment Change

We next evaluate the robustness of our system to the environment changes. Specifically, we introduce environment changes including furniture change and people walking around that described in the experimental setup when the participant is doing finger gestures. We then compare the performance of our system with and without using environmental noise removal technique. Figure 17 depicts the performance comparison for each gesture recognition in the home environment. We find that the environmental noise removal technique improves the performance significantly for each of the finger gesture under both furniture change and people walking around scenarios. Moreover, by comparing Figure 17 with Figure 12(a), we observe that the performance doesn't have obvious degradation due to the use of environmental noise removal technique. In addition, we find that people walking around has larger impact on the CSI measurements, as indicated by the performance under the case without using environmental noise removal. This study demonstrates that our system can effectively mitigate the impact from the surrounding objects or people and is robust to the environment changes.

### 4.5 Impact of Individual Diversity

We further test the resilience of our system to individual diversity by applying the gesture profile built from one participant to another participant. We compare the performance of our system to the one without using the gesture pattern extraction method. Figure 18 presents the performance comparison for each finger gesture under individual diversity. We observe that without using gesture pattern extraction method, the performance degrades dramatically due to individual diversity and gesture inconsistency. Our system, with the gesture pattern extraction, provides much higher recognition accuracy than that of without using gesture pattern extraction method. For example, our system could improve the recognition accuracy by over **10%** for most of the finger gestures. These results show that by incorporating the pattern extraction method, our system is resilient to individual diversity. Our system, once setup, could be used by multiple users without user-specific calibration.

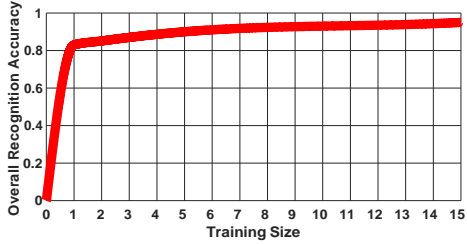

**Fig. 19: System performance under different training size.**

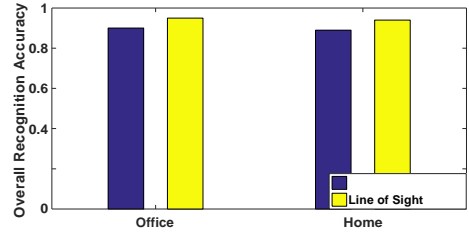

**Fig. 21: System performance under both NLOS and LOS scenarios.**

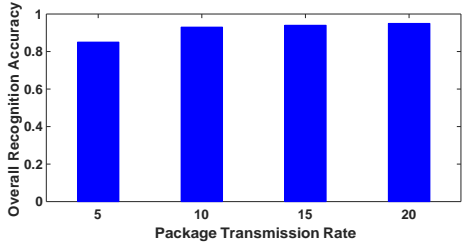

**Fig. 20: System performance under different packet rate.**

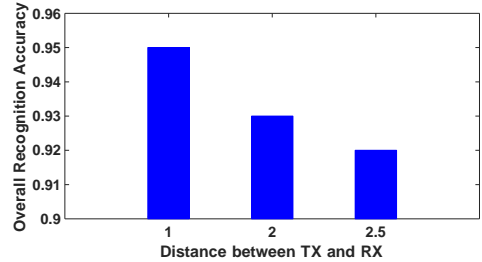

**Fig. 22: System performance under different distance between RX and TX.**

### 4.6 Impact of Training Size

When building the profile for each finger gesture, our system requires to extract the CSI patterns from multiple gesture instances. We refer such number as the training size. Figure 19 studies the impact of training size to the performance of our system. For this particular study, we focus on studying the single user scenario where one user is performing the finger gestures at a time. During the training phase, the profile is built upon data collected from both locations (i.e., home and office). Moreover, the data is selected from home location when the training size is one. During the evaluation phase, the profile is testing against the data collected from both home and office locations with different multi-path environments. Overall, we observe that our system could achieve considerable accuracy with a few tracing instances. In particular, with only one training size, our system achieves more than 80% of the recognition accuracy with one training instance. And the accuracy is improved to over 90% with five training instances. This result shows that our system could provide accurate gesture recognition with only a few training instances and hence doesn't incur high overhead on building the gesture profile, especially when the built profile from one user could be used by others.

### 4.7 Impact of Packet Rate

As a higher packet transmission rate results in more CSI measurements to capture the finger gestures, we are thus interested in whether existing WiFi traffic is sufficient to provide accurate gesture recognition. We experiment with four packet transmission rates, 5 pkts/s, 10 pkts/s, 15 pkts/s, and 20 pkts/s. The results are shown in Figure 20. We observe that a higher transmission rate results in a better recognition accuracy. Moreover, with 10 pkt/s transmission rate, our system is able to achieve more than 90% recognition accuracy. This demonstrates that our system could work with very low packet transmission rate. As the commercial AP sends beacon signals at 10 pkts/s, our system thus can reuse existing WiFi beacon signals for accurate gesture recognition.

### 4.8 Impact of NLOS

We study the impact of NLOS by placing the WiFi device and the AP in two connected rooms with the door closed. When the door is open, there exists LOS between the AP and the WiFi device. Figure 21 presents the performance comparison under the NLOS and LOS scenarios in both office and home environments. Results show that NLOS slightly degrades the system performance. Still, NLOS scenario has the overall recognition accuracy at around **90%** in both office and home environments. It demonstrates that the proposed system could even work under the NLOS scenario. This allows us to deploy the proposed system to a wider range of application domains.

### 4.9 Impact of TX-RX Distance

We study the impact of TX-RX distance by varying the distance between the WiFi device and the AP. We experiment with three different TX-RX distance, 1m, 2m, 2.5m. The results are shown in Figure 22. We observe that a closer distance results in a better overall recognition accuracy. Moreover, with 2.5m distance between TX-RX, our system is able to achieve around **92%** recognition accuracy. This demonstrates that our system could work under various TX-RX distance setups.

### 4.10 Impact of Threshold Value

We also study the impact of threshold value by applying different thresholds for sliding window in multi-user separation and similarity score in gesture identification. As shown in Figure 23, we observe that the overall recognition accuracy for multiple users across different gesture is around 92% when the appropriate thresholds are chosen for both sliding windows size and similarity score value. Furthermore, the accuracy is over 90% when sliding windows threshold value is between 0.58 to 0.67 and similarity score threshold value is between 0.73 and 0.8 respectively.

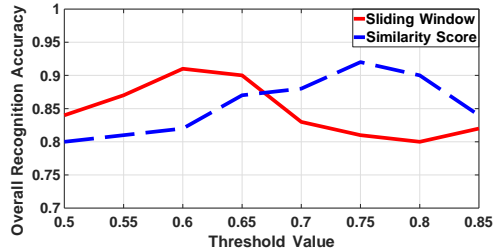

**Fig. 23: System performance under different threshold.**

## 5 Discussion

**Machine Learning Techniques.** Our system currently utilizes empirically selected thresholds to achieve higher overall recognition accuracy with existing data. Such an approach could affect system performance especially under the massive deployment scenarios. Hence, we propose to use more sophisticated machine learning methods (e.g., CNN, RNN, etc.) to determine various thresholds instead of empirically selection. For example, during the system setup phase, WiFinger can utilize initial data collections and machine learning algorithms to automatically determine the threshold values for multi-user profile separation, critical subcarrier selection and wavelet based denoising. Furthermore, those thresholds can be frequently updated when the new data is collected without user intervention. We would like to explore this as our future work to further improve system accuracy and robustness under different scenarios.

**Implementation Overhead.** The wavelet-based denoising technique is utilized to filter out environmental noises. Such a technique has a very low implementation overhead because it has been widely used in image processing, voice recognition, etc. The MD-DTW used for gesture identification is a comparably computational heavy approach, but the processing time depends on various factors including the number of enrolled finger gestures and the computational power of the system. Currently, our system runs on Dell Latitude E5540 laptop with dual-core CPU running at 1.9GHz, 4GB memory, and a built-in graphic processor. It takes an average of about 3 seconds to complete the recognition process. We believe that by leveraging widely available GPU based parallelization with higher computational power on the household computer, it is possible to cut the processing time to complete the recognition procedure under 1 second which is comparable to other gesture recognition systems.

**Effect of User Breathing.** The chest movements and the fine finger motions share a similar magnitude. But compare to fine finger motions (usually ranges from 1Hz to 4Hz), the chest movements have a much lower frequency (usually ranges from 0.2Hz to 0.5Hz). Thus, we can use a band-pass filter to mitigate the breath motion noises with lower frequency. We would like to include the study of chest motion on system performance as part of future work.

## 6 Conclusion

In this paper, we exploit the prevalence of WiFi devices and networks and design a system called WiFinger to perform fine-grained finger gesture recognition by utilizing the detailed CSI available in commodity WiFi devices. We find that CSI can capture the subtle movements of finger gestures. Our system benefits from such observation and examines the unique pattern exhibits in the detailed CSI for gesture recognition. To address the challenge of signal dynamic due to the environment changes, we devise environment noise removal mechanism to filter out the environmental noise while keeping the CSI pattern resulted from the finger gesture. Moreover, we propose to capture the intrinsic gesture behavior and to select critical subcarriers for accurate gesture recognition. Additionally, to achieve multi-user compatibility, we utilize multiple transmission links and the large bandwidth at 5GHz to separate the signal reflection from individual user. Extensive experiments in both home and office environments demonstrate that WiFinger is effective in distinguishing a number of finger gestures, and that it can achieve over 90% recognition accuracy. In addition, we show that our system can work with WiFi beacon signals, provide considerable recognition accuracy under NLOS scenario and work with multi-user scenario.